\documentclass[aps,prd,showpacs,amssymb,nofootinbib,twocolumn]{revtex4} 

\usepackage{amsmath}
\usepackage{amssymb}
\usepackage{latexsym}
\usepackage{amsfonts}
\usepackage{epsfig}
\usepackage{psfrag}
\usepackage{graphicx}

\usepackage{color}

\definecolor{black}{rgb}{0,0,0}
\definecolor{blue}{rgb}{0,0,1}
\definecolor{green}{rgb}{0,1,0}
\definecolor{red}{rgb}{1,0,0}
\definecolor{brown}{rgb}{0.4,0.2,0}
\definecolor{darkgreen}{rgb}{0,0.7,0}

\newcommand{\black}[1]{{\color{black}#1}}

\usepackage{bm}
\usepackage{amsmath}
\usepackage{amssymb}
\usepackage{latexsym}
\usepackage{amsfonts}
\usepackage{epsfig}
\usepackage{psfrag}
\usepackage{graphicx}

\newcommand{\nn}{\nonumber\\}

\newcommand{\bea}{\begin{eqnarray}}
\newcommand{\ea}{\end{eqnarray}}
\newcommand{\eea}{\end{eqnarray}}
\newcommand{\ord}{{\cal O}}

\begin{document}

\title{Hawking radiation with dispersion versus breakdown of WKB}

\author{R.~Sch\"utzhold$^1$}
\email[e-mail:\,]{ralf.schuetzhold@uni-due.de}

\author{W.~G.~Unruh$^2$}
\email[e-mail:\,]{unruh@physics.ubc.ca}

\affiliation{
$^1$Fakult\"at f\"ur Physik, Universit\"at Duisburg-Essen, 
Lotharstrasse 1, 47057 Duisburg, Germany
\\
$^2$CIAR Cosmology and Gravity Program, 
Dept.\ of Physics, University of B.C., Vancouver, Canada V6T 1Z1 
}

\date{\today}

\begin{abstract}
Inspired by the condensed matter analogues of black holes (a.k.a.~dumb holes),
we study Hawking radiation in the presence of a modified dispersion relation 
which becomes super-luminal at large wave-numbers. 
In the usual stationary coordinates $(t,x)$, one can describe the asymptotic 
evolution of the wave-packets in WKB, but this WKB approximation breaks down 
in the vicinity of the horizon, thereby allowing for a mixing between initial 
and final creation and annihilation operators.
Thus, one might be tempted to identify this point where WKB breaks down with 
the moment of particle creation.
However, using different coordinates $(\tau,U)$, we find that one can evolve 
the waves so that WKB in these coordinates is valid throughout  
this transition region -- which contradicts the above identification of the 
breakdown of WKB as the cause of the radiation. 
Instead, our analysis suggests that the tearing apart of the waves into two 
different asymptotic regions (inside and outside the horizon) is the major 
ingredient of Hawking radiation. 
\end{abstract}


\maketitle

\black{

\section{Introduction}

Despite decades of research, the precise nature of Hawking \cite{hawking}
radiation -- one of the most fundamental predictions of quantum field theory 
\cite{Birrell} in curved space-times -- is not fully understood yet.
E.g., open questions are: 
\begin{itemize}
\item
Why does Hawking radiation give a thermal spectrum, i.e., why do black holes 
seem to behave as thermal \cite{thermo} objects with an energy (mass), 
temperature, and because the temperature depends on energy, an entropy etc.?
\item 
What are the essential ingredients of Hawking radiation and how robust is it? 
Is the thermal spectrum and density matrix robust, or is it possible to encode 
some information into the seemingly thermal radiation given off?
\end{itemize}
The last point is especially relevant for the black hole information 
``paradox'', i.e., the question of whether (and if yes, how) the process of a 
gravitational collapse to a black hole and its subsequent evaporation can be 
described by a globally  unitary (i.e., information preserving)} process
\cite{Liberati-2010}.  
Because the field equations are local the system is always locally unitary. 
The question is essentially whether there are places (e.g., singularities, 
baby universes, etc.) into which information could be lost.

Sometimes it is argued that the information 
(possibly stored near the singularity or in the vicinity of the horizon) 
might perhaps leak out the black hole (during its evaporation process) 
hidden in correlations between the emitted Hawking quanta -- 
this would require some sort of imprinting procedure during which the 
information is stored (remembered) and later  transfered to the emitted 
particles. 
These motivate a better understanding of the robustness and origin of 
Hawking radiation.  
In the following, we study this question for a set-up including a modified 
dispersion relation  at large wave-numbers -- an approach which is partly
motivated by the condensed matter analogues \cite{unruh-prl}
of black holes (a.k.a.~dumb holes). 

\section{conventional derivation}

Let us start with a brief sketch of the conventional derivation of Hawking 
radiation in the presence of a modified dispersion relation 
\cite{universality}.
Usually, this is done in terms of the Painlev{\'e}-Gullstrand-Lema{\^\i}tre 
or the Eddington-Finkelstein coordinates \cite{PGL}. 
Here, we employ the former since they are more closely related to the 
condensed matter analogues of black holes.  
In terms of these coordinates, the metric reads ($\hbar=c=1$)
\bea
\label{PGL}
ds^2
&=&
[1-v^2(x)]dt^2-2v(x)dt\,dx-dx^2
\nn
&=&
dt^2-[dx+v(x)dt]^2
\,,
\ea
where we restrict our consideration to 1+1 dimensions for simplicity. 
Here $v(x)$ corresponds to the fluid velocity of the sonic black hole 
analogues and we have a horizon where $v(x_{\rm horizon})=c=1$.

In order to include a modified dispersion relation, the standard wave 
equation $\Box\phi=0$ is replaced by 
\bea
\label{modified}
\Box\phi=F(\partial_x)\phi
\,,
\ea
where $F$ contains higher order spatial derivatives. 
Here we use $F=-\gamma^2\partial_x^4$, which corresponds to the dispersion 
relation of phonons in atomic Bose-Einstein condensates \cite{garay}, 
for example, but our results can be generalized to other functions $F$. 

The scale $\gamma$ sets the cut-off wave-number $k_{\rm cutoff}=1/\gamma$ 
where the dispersion relation starts to deviate significantly from a linear 
behavior.
For atomic Bose-Einstein condensates, it is related to the healing 
length -- whereas for real gravity, one might speculate that it is 
related to the Planck scale. 
Then Eq.~(\ref{modified}) becomes 
\bea
\left([\partial_t-\partial_xv][\partial_t-v\partial_x]-
\partial_x^2+\gamma^2\partial_x^4
\right)\phi=0
\,.
\ea
The stationarity of this wave equation allows us to make the 
separation ansatz $\phi_{\omega}(x)e^{-i\omega t}$ and after 
a spatial WKB approximation, we get the dispersion relation  
\bea
\label{dispersion}
(\omega-vk)^2=k^2+\gamma^2 k^4
\leadsto
\omega=vk\pm\sqrt{k^2+\gamma^2 k^4}
\,,
\ea
where the square root on the r.h.s.\ is the dispersion 
relation $\omega(k)$ in Minkowski space-time (where $v=0$).
Without loss of generality, we set $x_{\rm horizon}=0$ 
such that the velocity $v(x)$ can be approximated in the vicinity 
of the horizon by 
\bea
\label{linearization}
v(x)=1-\kappa x+\ord(x^2)
\,,
\ea
where $\kappa$ is the surface gravity.
Assuming that the dispersion scale $1/\gamma$ is huge compared 
to all other scales, such as $\gamma\omega\lll1$, we may find 
an intermediate regime for $x$ 
\bea
\label{intermediate}
\gamma^2\omega^2\ll|\kappa x|^3\ll1
\,.
\ea
In this range, the linearization (\ref{linearization}) applies and
we find two WKB solutions with large and nearly opposite wave-numbers 
$k_\pm=\pm\sqrt{2|\kappa x|/\gamma^2}$ for $x<0$, i.e., 
inside the black hole. 
The associated WKB solutions of the wave equation read 
$\exp\{-i\omega t\pm\sqrt{8\kappa|x^3|/(9\gamma^2)}\}$.
There are also other solutions of the dispersion relation (\ref{dispersion}),  
but they have much smaller $k$-values $k=\ord(\omega)$ and are thus well 
separated from the $k_\pm$ solutions. 

However, when $\kappa x$ becomes too small (i.e., very close to the horizon) 
such that $|\kappa x|^3\sim\gamma^2\kappa^2$, this WKB approximation breaks 
down.   
One way to see this is to take the spatial derivative of the 
dispersion relation (\ref{dispersion}) giving 
\bea
\label{derivative}
\frac{dk}{dx}=-\frac{dv}{dx}\,\frac{k}{v(x)\pm v_{\rm group}(k)}
\,,
\ea
where $v_{\rm group}$ is group velocity $d\omega/dk$ in Minkowski space-time
(where $v=0$). 
When approaching the horizon (where $v=1$), $dk/dx$ diverges and thus the 
WKB approximation breaks down.
At that point, two real $k$-solutions of the dispersion relation 
(\ref{dispersion}) merge and become complex -- i.e., the modes do 
not stay separated and mix. 
Thus, one might be tempted to identify this point with the place of 
particle creation. 

\section{linear profile} 

Since the critical point where the WKB approximation described above breaks 
down is very close to the horizon $|\kappa x|^3\sim\gamma^2\kappa^2\lll1$, 
we focus on this region and employ the near-horizon approximation by setting 
\bea
\label{linear}
v(x)=1-\kappa x
\,.
\ea
Introducing the usual Kruskal light-cone coordinate $U$,  
the metric (\ref{PGL}) can then be cast into the form 
\bea
U=-x\exp\left\{-\kappa t\right\}
\,\leadsto\,
ds^2=2e^{\kappa t}dt\,dU-e^{2\kappa t}dU^2
\,.
\ea
With $\sqrt{-g}=e^{\kappa t}$, we get the d'Alembert operator 
\bea
\Box\phi
=
e^{-\kappa t}
\left(
\partial_te^{\kappa t}\partial_t+2\partial_U\partial_t
\right)\phi
\,,
\ea
and the wave equation (\ref{modified}) with a modified dispersion relation
$F=-\gamma^2\partial_x^4$ reads 
(note that $\partial_x=-e^{-\kappa t}\partial_U$) 
\bea
\left(
\partial_te^{\kappa t}\partial_t+2\partial_U\partial_t
\right)\phi
=-\gamma^2 e^{-3\kappa t}\partial_U^4\phi
\,.
\ea
Now $U$ corresponds to a Killing vector (but not $t$), 
and we can make the separation ansatz $\phi(t,U)=\phi_K(t)e^{iKU}$
\bea
\ddot\phi_K+\kappa\dot\phi_K+2iKe^{-\kappa t}\dot\phi_K
=
-\gamma^2 K^4e^{-4\kappa t}\phi_K
\,.
\ea
\black{For late times $t\uparrow\infty$, the exponential factors 
$e^{-\kappa t}$ and $e^{-4\kappa t}$ can be neglected and thus the 
above equation corresponds to the motion of a particle with damping.
Hence} we see that the solution freezes 
$\phi(t\uparrow\infty,U)=\phi_K^\infty e^{iKU}$
at late times $t\uparrow\infty$. 
This motivates the introduction of a new time coordinate via 
\bea
\tau=-\frac{1}{\kappa}\,e^{-\kappa t}
\,\leadsto\,
d\tau=e^{-\kappa t}dt
\,,
\ea
such that the wave equations simplifies to 
\bea
\left(\partial_\tau^2+2iK\partial_\tau\right)\phi_K
=
-\gamma^2 K^4(\kappa\tau)^2\phi_K
\,.
\ea
With the re-definition $\varphi_K(\tau)=e^{iK\tau}\phi_K(\tau)$, 
we may eliminate the $2iK\partial_\tau$ term and obtain the simple form 
\bea
\label{Whittaker}
\left(\partial_\tau^2+\gamma^2 K^4(\kappa\tau)^2+K^2\right)\varphi_K=0
\,.
\ea
Equations of this type are well-known and can be solved in terms of 
Whittaker or Kummer functions \cite{Abramowitz-Stegun}. 
Incidentally, the same equation occurs for the Sauter-Schwinger effect,
i.e., electron-positron pair creation out of the QED vacuum due to a 
strong electric field \cite{sauter}. 

\section{WKB analysis -- linear profile}\label{WKB-linear}

Even though the wave equation (\ref{Whittaker}) can be solved exactly, 
let us find an approximate solution via the WKB method.
The associated dispersion relation reads 
\bea
\label{dispersion-tau}
\Omega_K(\tau)=\pm K
\sqrt{1+\gamma^2\kappa^2(K\tau)^2}
\,.
\ea
If we now check the applicability of the WKB approximation in analogy to 
Eq.~(\ref{derivative}), we find that   
\bea
\gamma\kappa\ll1
\,\leadsto\,
\left|
\frac{1}{\Omega^2_K}
\frac{d\Omega_K}{d\tau}
\right|
\ll1
\,,
\ea
and thus the WKB method applies for all $\tau$, i.e.,  
all the way from $t\downarrow-\infty$ down to $t\uparrow+\infty$.
The two branches of solutions of (\ref{dispersion-tau}) with positive 
and negative $\Omega_K(\tau)$ stay well separated throughout the 
evolution and there is negligible mode mixing between them.
 
Let us study the wave-packet trajectories for the two branches.
First, we consider the case without dispersion $\gamma=0$ for comparison. 
For the branch with positive $\Omega_K(\tau)$, the solution behaves as 
$e^{iKU}$ and thus the trajectories have $U=\rm const$, 
i.e., $x\propto e^{\kappa t}$.
These are the modes peeling off the black hole horizon at $U=0$ 
(i.e., $x=0$), for example the Hawking modes.
For the other branch with negative $\Omega_K(\tau)$, the solution behaves 
as $e^{iKU-2iK\tau}$ and thus the trajectories have $U=2\tau+\rm const$,
i.e., $x=2/\kappa+e^{\kappa t}\times\rm const$.
At $x=2/\kappa$, there is a white hole horizon (where $v=-1$) according to 
(\ref{linear}).  
Thus, the modes corresponding to the other branch peel off the white hole 
horizon at $x=2/\kappa$ but they propagate freely across the black hole 
horizon.  

As one would expect, this behavior changes in the presence of a modified 
dispersion relation (\ref{dispersion-tau}).
In this case, the modes which later form the Hawking radiation do not 
originate from the vicinity of the horizon at $U=0$ but approach the horizon 
from the inside in the past. 
In analogy to (\ref{intermediate}), we use the smallness of 
$\gamma\kappa\lll1$ to find an intermediate regime for $\tau$ where 
\bea
\label{intermediate-tau}
\frac{1}{\gamma^2\kappa^2}\gg(K\tau)^2\gg1
\,.
\ea
In this regime, we may Taylor expand the square root in (\ref{dispersion-tau})
and thus the final Hawking modes initially behave as 
$\exp\{iKU+i\gamma^2\kappa^2K^3\tau^3/6\}$. 
Using the stationary phase (or saddle point) approximation, we find the WKB  
trajectories as $U=-\gamma^2\kappa^2K^2\tau^3/2$.
As expected, the modes approach the horizon from the inside where $U>0$
and $x<0$ (note that $\tau<0$). 
The above condition (\ref{intermediate-tau}) also ensures that 
$\kappa x=U/\tau=-\gamma^2\kappa^2K^2\tau^2/2$ is small 
and thus consistent with our linear approximation (\ref{linear}).  

\section{particle creation}

Now, after having studied the solutions of the wave equation, 
let us derive the consequences for Hawking radiation. 
To this end, we use the same trick as in \cite{unruh}
and consider the following function 
\bea
\phi_\omega(U)=
\left\{
\begin{array}{lll}
e^{+\pi\omega/(2\kappa)}\,\left|\kappa\,U\right|^{i\omega/\kappa} 
& {\rm for} & U>0 \\
e^{-\pi\omega/(2\kappa)}\,\left|\kappa\,U\right|^{i\omega/\kappa} 
& {\rm for} & U<0 
\end{array}
\right.
\,.
\ea
After analytic continuation, this function is holomorphic in the entire 
upper half of the complex $U$ plane $\Im(U)>0$ 
while it has a branch cut in the lower half.
As a result, the Fourier decomposition of this function $\phi_\omega(U)$
consists of modes $e^{iKU}$ with positive $K>0$ only.
(For negative $K<0$, the integral determining the overlap with 
$\phi_\omega(U)$ can be closed in the upper half of the complex plane
and thus vanishes). 

As we have seen in the previous Section, the modes which finally behave 
as $\exp\{iKU\}$ are initially of the form 
$\exp\{iKU+i\gamma^2\kappa^2K^3\tau^3/6\}$. 
In terms of the original $x$-coordinate, we get 
$\exp\{iKx\kappa\tau+i\gamma^2\kappa^2K^3\tau^3/6\}$.
Now, evaluating the time-derivative for fixed $x$, we find that these 
wave-packets are slowly varying along their trajectories 
$\kappa x=U/\tau=-\gamma^2\kappa^2K^2\tau^2/2$.  
This is consistent with the fact that these wave-packets will later 
transform into the modes constituting the Hawking radiation -- which 
has a small frequency $\omega=\ord(\kappa)$ that is conserved. 

As the next step, let us study how the initial mode of the form 
$\exp\{iKx\kappa\tau+i\gamma^2\kappa^2K^3\tau^3/6\}$ is experienced 
by a freely falling observer with the trajectory 
$x(t)\approx x_0-t$.  
Since (as we have seen above), the temporal variance is slow compared 
to the spatial variation, the freely falling observer will see a rapid 
oscillation with a frequency of $\omega_{\rm in}\approx K\kappa\tau$ 
initially, i.e., for $(K\tau)^2\gg1$.
Thus, modes with positive $K$ will have a large negative frequency 
$\omega_{\rm in}$ due to $\tau<0$ and vice versa. 

As the final ingredient, let us convert the above function $\phi_\omega(U)$ 
into $t,x$ coordinates, where we get  
\bea
\label{convert}
\phi_\omega(t,x)=
\left\{
\begin{array}{lll}
e^{+\pi\omega/(2\kappa)}\,
\left|\kappa\,x\right|^{i\omega/\kappa}\,e^{-i\omega t}
& {\rm for} & x<0 \\
e^{-\pi\omega/(2\kappa)}\,
\left|\kappa\,x\right|^{i\omega/\kappa}\,e^{-i\omega t}
& {\rm for} & x>0 
\end{array}
\right.
\,.
\ea
We see that we obtain the wave-functions of the outgoing Hawking radiation 
(with positive pseudo-norm) for $x>0$ and its in-falling partner particles 
(with negative pseudo-norm) for $x<0$ 
(where both have the same conserved frequency $\omega$). 
Therefore, if we consider the evolution of a Gaussian wave-packet 
peaked at $K_0\gg\Delta K>0$
\bea
\phi_{K_0,\Delta K}(\tau,U)
\propto
\int dK\,\phi_K(\tau)
e^{iKU-(K-K_0)^2/\Delta K^2}
\,,
\ea
we find the following picture:
Initially, i.e., for values of $\tau$ satisfying (\ref{intermediate-tau}),
this wave-packet approaches the horizon from the inside according to 
the trajectory $U=-\gamma^2\kappa^2K^2_0\tau^3/2$.
For a freely falling observer, it has a large negative frequency 
$\omega_{\rm in}\approx K_0\kappa\tau$.
During the approach to the horizon, the wave-packet is constantly 
red-shifted according to $\omega_{\rm in}\approx -K_0e^{-\kappa t}$.
This reflects the time-translation invariance of our system: 
Modes with larger $K_0$ evolve in the same way as those with smaller 
$K_0$, but at later times $t$.
Finally, this wave-packet is ripped apart at the horizon into an 
outgoing part and its in-falling partner. 
Nevertheless, the WKB method in Sec.~\ref{WKB-linear} 
remains applicable during the whole process. 
For a given final frequency $\omega$, the amplitudes of these two parts 
directly yield the Bogoliubov coefficients $\alpha_\omega$ and $\beta_\omega$
and from (\ref{convert}), we may read off 
\bea
\left|\frac{\beta_\omega}{\alpha_\omega}\right|
=
\exp\left\{-\pi\,\frac{\omega}{\kappa}\right\}
\,.
\ea
Together with the normalization $|\alpha_\omega|^2-|\beta_\omega|^2=1$,
this gives a thermal spectrum 
\bea
\langle\hat n_\omega^{\rm out}\rangle_{\rm in}
=
|\beta_\omega|^2
=
\frac{1}{\exp\{2\pi\omega/\kappa\}-1}
\,,
\ea
with the Hawking temperature ($\hbar=c=k_{\rm B}=1$)
\bea
T_{\rm Hawking}=\frac{\kappa}{2\pi}
\,.
\ea
%

\section{general profile}

For a more general velocity profile $v(x)$, the light cones lie at 
$dt=\pm[dx+v(x)dt]$ and thus the Kruskal light-cone variable $U$ reads 
\bea
U=-\frac{1}{\kappa}\exp\left\{-\kappa t-\kappa\int\frac{dx}{v(x)-1}\right\}
\,,
\ea
where $\kappa=|v'(x_{\rm horizon})|$ is again the surface gravity. 
Inserting the coordinate differential 
\bea
dU=-\kappa U\left(dt+\frac{dx}{v(x)-1}\right)
\,,
\ea
the metric becomes
\bea
ds^2=2\,\frac{v-1}{\kappa U}\,dU\,dt-\left(\frac{v-1}{\kappa U}\right)^2dU^2
\,.
\ea
Note that the factor 
\bea
\chi=\sqrt{-g}=\frac{v-1}{\kappa U}
\ea
is regular at the horizon and behaves as $\chi=e^{\kappa t}f(x)$
with a regular function $f(x)$ satisfying $f(x_{\rm horizon})=1$. 
Using the same modification $F(\partial_x)$ of the dispersion relation as 
before, the wave equation becomes 
\bea
\label{wave-equation-chi}
\Box\phi
=
\frac{1}{\chi}\left(\partial_t\chi\partial_t+2\partial_U\partial_t\right)\phi
=
F(\partial_x)\phi=-\gamma^2\partial_x^4\phi
\,,
\ea
where $\partial_x=-\chi^{-1}\partial_U$. 

\section{tanh-profile}

In order to deal with a smooth profile with well-defined asymptotics, 
let us consider the following example 
\bea
v(x)=1-\tanh(\kappa x)
\,.
\ea
In this case, the Kruskal variable $U$ simply becomes 
\bea
U=-\frac{\sinh(\kappa x)}{\kappa}\exp\left\{-\kappa t\right\}
=\tau\sinh(\kappa x)
\,,
\ea
and the factor $\chi$ in $ds^2=2\chi dU dt-\chi^2dU^2$ reads 
\bea
\chi
=
\frac{e^{\kappa t}}{\cosh(\kappa x)}
=
\frac{1}{\sqrt{e^{-2\kappa t}+\kappa^2U^2}}
=
\frac{\kappa^{-1}}{\sqrt{\tau^2+U^2}}
\,.
\ea
The wave equation (\ref{wave-equation-chi}) assumes the form 
\bea
\label{wave-equation-tanh}
\left(
\chi\partial_t\chi\partial_t+2\chi\partial_t\partial_U
+\gamma^2\chi^2[\chi^{-1}\partial_U]^4
\right)\phi=0
\,,
\ea
and, after transforming to the time coordinate $\tau$, we get 
\bea
\left(
\xi\partial_\tau\xi\partial_\tau+2\xi\partial_\tau\partial_U
+\gamma^2\chi^2[\chi^{-1}\partial_U]^4
\right)\phi=0
\,,
\ea
with the purely $x$-dependent factor 
\bea
\xi
=
\frac{1}{\cosh(\kappa x)}
=
\frac{1}{\sqrt{1+U^2/\tau^2}}
\,.
\ea
New let us consider the two limiting cases.
For $\tau^2\gg U^2$ (i.e., small $x$), we have $\xi\to1$ and thus 
the wave equation simplifies and we get the same result as with 
the linear profile (as expected) 
\bea
\left(
\partial_\tau^2+2\partial_\tau\partial_U
+\gamma^2\kappa^2\tau^2\partial_U^4
\right)\phi=0
\,.
\ea
In the other limiting case $\tau^2\ll U^2$ (i.e., large $x$),
we have $\chi\to1/|\kappa U|$ and thus wave equation becomes 
approximately independent of $t$ 
\bea
\left(
\partial_t^2+2|\kappa U|\partial_t\partial_U
+\gamma^2[\kappa U\partial_U]^4
\right)\phi=0
\,.
\ea
This equation has two branches of solutions 
\bea
\phi^\pm_\theta
\propto
|\kappa U|^{i\theta}
\exp\left\{
-i\kappa\theta t\left(1\pm\sqrt{1+\gamma^2\kappa^2\theta^2}\right)
\right\}
\,,
\ea
labeled by $\theta=\rm const$. 
Without dispersion, i.e., for $\gamma=0$, we may interpret the two branches 
in the same way as in Section~\ref{WKB-linear}: 
one $\phi^+\propto|\kappa U|^{i\theta} e^{-2i\kappa\theta t}$ 
is propagating into the black hole while the other 
$\phi^-\propto|\kappa U|^{i\theta}$ is trying to escape.  
However, most interesting is the case with dispersion, where we get 
\bea
\phi^-_\theta\sim
\exp\left\{
i\kappa\theta\left(|x|-2t+t\sqrt{1+\gamma^2\kappa^2\theta^2}
\right)
\right\} 
\,.
\ea
We see that the modes with $\gamma^2\kappa^2\theta^2<3$ are propagating 
away from the horizon while the others with $\gamma^2\kappa^2\theta^2>3$
are moving towards it. 
A final Hawking particle with frequency $\omega>0$ stems from initial 
modes with $\theta<0$ and 
\bea
\gamma^2\kappa^2\theta^2=3+\sqrt{\frac{16\gamma^2\omega^2}{3}}
\,,
\ea
which propagate very slowly towards the horizon. 
Again, for a freely falling observer, these modes with 
$\theta<0$ and $|\theta|\approx\sqrt{3/(\gamma^2\kappa^2)}\gg1$ 
have a very large negative frequency. 


\section{WKB analysis -- tanh-profile}

Since we can solve the wave equation (\ref{wave-equation-tanh}) exactly in 
the two limiting cases $\tau^2\gg U^2$ and $\tau^2\ll U^2$, the remaining 
critical regime is where $U$ and $\tau$ are roughly of the same order, 
i.e., where $\kappa x=\ord(1)$.  
In this regime, we may obtain $\Omega_K$ and $K$ via the transformation 
from the original $t,x$ coordinates to the $\tau,U$ coordinates
\bea
\left(
\begin{array}{c}
\partial_t \\
\partial_x
\end{array}
\right)
=
-\kappa
\left(
\begin{array}{cc}
\tau & U \\
0 & \sqrt{U^2+\tau^2} 
\end{array}
\right)
\cdot
\left(
\begin{array}{c}
\partial_\tau \\
\partial_U
\end{array}
\right)
\,.
\ea
Identifying $\partial_t\to\Omega$ and $\partial_x\to k$, as well as,  
$\partial_\tau\to\Omega_K$ and $\partial_U\to K$, we get the transformation  
\bea
K=-\frac{k}{\sqrt{\kappa^2U^2+\kappa^2\tau^2}}=-k\chi
\,.
\ea
Furthermore, we find the exact conservation law 
\bea
\tau\Omega_K+UK=-\frac{\omega}{\kappa}=\rm const
\,,
\ea
which reflects the fact that $\partial_t$ is a Killing vector in the 
original $t,x$ coordinates and thus the solution behaves as 
$e^{i\omega t+iS(x)}$ with an eikonal function $S(x)$ depending on 
$\kappa x={\rm arsinh}(U/\tau)$.
Finally, using that $k\gg\omega$ in the regime of interest, we find 
\bea 
\Omega_K
\approx 
-\frac{U}{\tau}\,K
=
\frac{k}{\sqrt{\kappa^2U^2+\kappa^2\tau^2}}\,\frac{U}{\tau}
\,.
\ea
Now, if $U$ and $\tau$ are roughly of the same order, i.e., 
for $\kappa x=\ord(1)$, we know that $k$ is large $k\gg\kappa$ 
and slowly varying (while $\omega$ is exactly constant).
Thus we find that $K$ and $\Omega_K$ are also slowly varying, 
for example   
\bea
\frac{1}{K^2}\,\frac{\partial K}{\partial U}
\approx
\frac{\kappa}{k}\,\frac{U}{\sqrt{U^2+\tau^2}}\leq\frac{\kappa}{k}\ll1
\,,
\ea
and similarly for the other terms.
Consequently, we again find that the WKB approximation does not breaks 
down when approaching the horizon. 

\section{conclusions}

In summary, we studied the origin of Hawking radiation in the presence of 
a modified dispersion relation at large $k$ which provides an effective  
UV cut-off.
We find that the transformation form $(t,x)$ coordinates to $(\tau,U)$ 
coordinates offers several advantages:
First, if the UV cut-off scale $1/\gamma$ is much larger than the surface 
gravity $\kappa$ (i.e., Hawking temperature), then the WKB approximation
remains valid throughout the evolution.
Second, the derivation of Hawking radiation becomes much simpler -- 
for example, we avoid dealing with non-trivial complex contour integrals 
involving branch cuts and saddle points, see, e.g., \cite{universality}.  

As a result of the validity of the WKB approximation throughout the evolution,
we find that the genesis of Hawking radiation is an extremely robust process 
-- as long as the (yet unknown) theory of quantum gravity incorporates 
some sort of general covariance, which allows us to go from $(t,x)$ 
coordinates to $(\tau,U)$ coordinates.
In view of the black hole information ``paradox'' mentioned in the 
Introduction, this robustness show that it is not obvious how to encode 
information in the outgoing Hawking photons. 

It is important to note that the thermal properties of the outgoing 
radiation does not come about because of any interaction with other 
degrees of freedom, a la Planck \cite{Planck}.
They are a direct consequence of the free field evolution. 
The particles have no obvious sources except for the tidal disruption of the 
evolution by the stretching of the wavelengths. 
The energy is locally conserved by the equations of motion, and does arise  
due to emission or absorption of that radiation. 

\section*{Acknowledgements}

The authors thank the Perimeter Institute for Theoretical Physics 
(Waterloo, Canada) for the kind hospitality and support for a research 
stay during which part of this work was done. 
W.G.U.~thanks both the NSERC of Canada for research support and the CIfAR 
for additional support during this work.
R.S.~acknowledges support from DFG (SCHU 1557/1-3, SFB-TR 12). 



\begin{thebibliography}{499}

\bibitem{hawking}
S.~W.~Hawking,
Nature {\bf 248}, 30 (1974);
Commun.\ Math.\ Phys.\ {\bf 43}, 199 (1975).

\bibitem{Birrell}
N.~D.~Birrell and P.~C.~W.~Davies,
{\em Quantum Fields in Curved Space},
(Cambridge University Press, Cambridge, England 1982).  
%

\bibitem{thermo}
J.~D.~Bekenstein,
Lett.\ Nuovo Cim.\ {\bf 4}, 737 (1972);
Phys.\ Rev.\ D {\bf 7}, 2333 (1973);
ibid.\ {\bf 9}, 3292 (1974);
ibid.\ {\bf 12}, 3077 (1975);
%
J.~M.~Bardeen, B.~Carter and S.~W.~Hawking,
Commun.\ Math.\ Phys.\ {\bf 31}, 161 (1973).

\bibitem{Liberati-2010} 
See, e.g., 
S.~Liberati, L.~Sindoni, S.~Sonego,
Gen.\ Rel.\ Grav.\ {\bf 42}, 1139 (2010);
and references therein. 

\bibitem{unruh-prl}
W.~G.~Unruh,
Phys.\ Rev.\ Lett.\ {\bf 46}, 1351 (1981); 
%
M.~Novello, M.~Visser, and G.~Volovik (eds.),
{\em Artificial Black Holes}
(World Scientific, Singapore, 2002);
%
G.~E.~Volovik,
{\em Universe in a Helium Droplet}
(Oxford University Press, Oxford, 2003);
%
C.~Barcel\'o, S.~Liberati, and M.~Visser,
Living Rev.\ Rel.\ {\bf 8}, 12 (2005);
%
R.~Sch\"utzhold and W.~G.~Unruh (eds.),
{\em Quantum Analogues: From Phase Transitions to Black Holes
\& Cosmology},
Springer Lecture Notes in Physics {\bf 718} (2007);
%
R.~Sch\"utzhold,
%
{Class.\ Quant.\ Grav.} {\bf 25}, 114011 (2008). 
%

\bibitem{universality}
See, e.g., 
T.~Jacobson,
Phys.\ Rev.\ D {\bf 44}, 1731 (1991);
%
R.~Brout, S.~Massar, R.~Parentani and P.~Spindel,
ibid.\ {\bf 52}, 4559 (1995);
%
S.~Corley,
ibid.\ {\bf 57}, 6280 (1998);
%
T.~Jacobson and D.~Mattingly,
ibid.\ {\bf 61}, 024017 (2000);
%
Y.~Himemoto and T.~Tanaka,
ibid.\ {\bf 61}, 064004 (2000);
%
H.~Saida and M.~Sakagami,
ibid.\ {\bf 61}, 084023 (2000).
%
W.~G.~Unruh and R.~Sch\"utzhold,
ibid.\ {\bf 71}, 024028 (2005);
%
R.~Sch\"utzhold and W.~G.~Unruh,  
ibid.\ {\bf 78}, 041504(R) (2008);
%
ibid.\ {\bf 81}, 124033 (2010);
%
A.~Coutant, R.~Parentani, S.~Finazzi,  
ibid.\ {\bf 85}, 024021 (2012);
%
U.~Leonhardt, S.~Robertson, 
New J.\ Phys.\ {\bf 14}, 053003 (2012);
%
S.~Robertson, 
J.\ Phys.\ B 
{\bf 45}, 163001 (2012). 

\bibitem{PGL}
P.~Painlev{\'e}, 
C.\ R.\ Hebd.\ Seances Acad.\ Sci.\ (Paris) {\bf 173}, 677 (1921);
%
A.~Gullstrand, 
Ark.\ Mat.\ Astron.\ Fys.\ {\bf 16}, 1 (1922);
%
G.~Lema{\^\i}tre, 
Ann.\ Soc.\ Sci.\ (Bruxelles) A {\bf 53}, 51 (1933);
%
A.~S.~Eddington,
Nature {\bf 113}, 192 (1924);
%
D.~Finkelstein,
Phys.\ Rev.\ {\bf 110}, 965 (1958).

\bibitem{garay}
L.~J.~Garay, J.~R.~Anglin, J.~I.~Cirac, and P.~Zoller,
Phys.\ Rev.\ Lett.\ {\bf 85}, 4643 (2000). 

\bibitem{Abramowitz-Stegun}
M.~Abramowitz, I.A.~Stegun (eds.), 
{\em Handbook of Mathematical Functions with Formulas, 
Graphs, and Mathematical Tables} (Dover, New York, 1972). 

\bibitem{sauter}
F.~Sauter,
Z.\ Phys.\  {\bf 69}, 742 (1931);
%
W. Heisenberg and H. Euler,
Z.\ Phys.\ {\bf 98}, 714 (1936);
%
J.~Schwinger, 
Phys.\ Rev.\ {\bf 82}, 664 (1951).

\bibitem{unruh}
W.~G.~Unruh,
Phys.\ Rev.\ D {\bf 14}, 870 (1976).

\bibitem{Planck}
M.~Planck, 
Verhandl.\ der Deutschen Physikal.\ Gesellsch.\ {\bf 2}, 202 (1900);
ibid.\ {\bf 2}, 237 (1900);
%
Ann.\ d.\ Phys.\ {\bf 306}, 69 (1900); 
ibid.\ {\bf 306}, 719 (1900). 

\end{thebibliography}
\end{document}